\documentclass[preprint2]{aastex6}
\usepackage{amssymb,amsmath}
\usepackage{multirow}
\usepackage{colortbl}
\usepackage{longtable}

\slugcomment{Accepted to the Astronomical Journal: January 21, 2017}

\begin{document}

\title{The structure of the distant Kuiper belt in a Nice model scenario}

\author{R. E. Pike\altaffilmark{1,2}}
\author{S. Lawler\altaffilmark{3}}
\author{R. Brasser\altaffilmark{4}}
\author{C. J. Shankman\altaffilmark{1}}
\author{M. Alexandersen\altaffilmark{2}}
\author{J.  J.  Kavelaars\altaffilmark{1,3}}

\altaffiltext{1}{Department of Physics and Astronomy, University of Victoria, Victoria, BC, Canada}
\altaffiltext{2}{Institute of Astronomy and Astrophysics, Academia Sinica, Taipei 10617, Taiwan}
\altaffiltext{3}{National Research Council of Canada, Victoria, BC, Canada}
\altaffiltext{4}{Earth Life Science Institute, Tokyo Institute of Technology, Meguro, Tokyo 152-8550, Japan}

\begin{abstract}
This work explores the orbital distribution of minor bodies in the outer Solar System emplaced as a result of a Nice model migration from the simulations of \cite{brasser2013}.
This planetary migration scatters a planetesimal disk from between 29-34 AU and emplaces a population of objects into the Kuiper belt region.
From the 2:1 Neptune resonance and outward, the test particles analyzed populate the outer resonances with orbital distributions consistent with trans-Neptunian object (TNO) detections in semi-major axis, inclination, and eccentricity, while capture into the closest resonances is too efficient.
The relative populations of the simulated scattering objects and resonant objects in the 3:1 and 4:1 resonances are also consistent with observed populations based on debiased TNO surveys, but the 5:1 resonance is severely underpopulated compared to population estimates from survey results.
Scattering emplacement results in the expected orbital distribution for the majority of the TNO populations, however the origin of the large observed population in the 5:1 resonance remains unexplained.

\end{abstract}

\section{Introduction}

The trans-Neptunian objects (TNOs) populate the region beyond Neptune, and the specifics of their formation location and evolutionary history are the subject of much study.
Early ideas of a quiescent belt, surviving beyond the giant planets, ware clearly incomplete based on the orbital characteristics of the early TNO discoveries \citep{nice}.
Even the first known TNO, Pluto, has large eccentricity and inclination; the TNOs must have been dynamically evolved in the past \citep{malhotra}.
The TNO population has dynamically excited eccentricity and inclination distributions, and the objects extend out to large semi-major axes.
In addition, the resonant populations are much larger than expected for a Kuiper belt which experienced no dynamical sweeping or scattering.
The over-population of objects in resonance is an indicator of a previous dynamical instability affecting the outer Solar System \citep{malhotra1993,malhotra,hahnmalhotra05,gomes2005,nice}.

The relative sizes and physical properties of the sub-populations in the Kuiper belt hold clues to the region's evolutionary history.
The sizes of the 3:2 and 2:1 resonant populations compared to the classical Kuiper belt and the relative sizes of the cold and hot classical populations are dependent on the specifics of the dynamical evolution \citep[e.g.][]{malhotra, chiang2002,nesvorny2015a}.
Some regions of the Kuiper belt have different physical properties; for example, the cold classical objects, with a less excited inclination distribution, have a steeper size distribution than the hot classical objects \citep{bernstein04}.
The surface colors of TNOs are also correlated with their dynamics; different color distributions correspond to different dynamical sub-populations \citep[e.g.][]{tegler03}.
\cite{brown} suggests that the surface colors of TNOs could be produced by forming these objects interior to their present location and moving them outward to their current orbits.
The outer Solar System population characteristics are complex and provide clues about the formation and evolution of the Solar System.

An acceptable model of Solar System evolution will reproduce these aspects of the TNO population.
A dynamical instability of the giant planets can increase both the eccentricity and inclination of small bodies and scatter a large number of TNOs into resonance \citep{nice}.
A slow migration of Neptune outward would capture TNOs into mean motion resonances \citep{malhotra}.
This slow sweeping pumps up the eccentricity of captured TNOs without significantly altering their pericenter distances.
The large binary fraction of some subpopulations of TNOs favors a slow or minimal migration because these binaries are likely to have been disrupted by more violent scattering interactions \citep{parker2010}.
Smooth migration scenarios typically result in TNO populations which are not sufficiently dynamically excited in inclination, and the possibility of more granular and less smooth migration models have been explored \citep{hahnmalhotra05,nesvorny2015a}.

\cite{thommes1999} suggested that the early configuration of giant planets was dynamically unstable.
The early incarnations of rapid planetary migration models are known as the `Nice model' \citep{tsiganis2005,morbidelli2005,gomes2005}.
In these scenarios, there is a large dynamical instability, such as Saturn and Jupiter crossing their 2:1 mean motion resonance, scattering Uranus and Neptune, which subsequently scatter small bodies, emplacing the TNOs and Oort cloud and causing the Late Heavy Bombardment.
This scenario also results in capture of the Jovian Trojan asteroids \citep{morbidelli2005}.
This more rapid planetary scattering event results in different characteristics of captured objects as compared to a smooth migration model.

A detailed comparison of TNO detections and numerical simulation results requires a carefully observed Kuiper belt in addition to a well-sampled simulation.
Several recent surveys of the Kuiper belt have attempted to provide TNO discoveries with known discovery biases \citep{schwamb2010, cfeps, adams14, alexandersen, bannister2016, hilat2016}.
These surveys characterize their discovery biases to facilitate comparison with population models.

This work explores the specific effects of the `Nice' model scenario on the scattering and resonant TNO populations in detail.
As a generic exploration of outer Solar System evolution, the predicted TNO populations from the Nice migration simulation by \cite{brasser2013} are tested against the real TNO detections from \cite{cfeps, hilat2016} and \cite{alexandersen}.
The combination of real detections and characterizations from the surveys and a survey simulator provides a powerful tool for comparing an external model to the survey detections.
The simulation and test particle classification are discussed in Section \ref{migration}.
An explanation of the survey simulator debiasing procedure is provided in Section \ref{comparing_section}.
Section \ref{classifying} presents the results of the classification and analysis; the B\&M model of the Kuiper belt, as a proxy for the scattering that likely occurs in a Nice model type scenario, does a reasonable job of populating the outer Solar System scattering components and resonances beyond $a\sim45$.
The notable exception is the 5:1 population, which is not well reproduced in this scenario, independent of the results from \cite{pike2015}.
The discussion and conclusions are presented in Section \ref{conclusions}.

\section{Migration Model}
\label{migration}

The model from \cite{brasser2013}\footnote{The B\&M model end state including the particle classifications for all test particles from $20<a<160$ AU, as generated in this work, is available at:  doi:10.11570/16.0009.} is examined here, referred to as the B\&M simulation.
The B\&M simulation is considered as an example of a Nice model type scenario and used to assess the accuracy of this model in producing the scattering and resonant TNO populations.
The TNO comparison population is from the Canada-France Ecliptic Plane Survey \citep[CFEPS;][]{cfeps,hilat2016} and the \cite{alexandersen} survey.
Using a dynamical simulated model from a source external to the observational survey results provides a useful test of the orbital distribution models created by the survey team.
This work focuses on the `scattered disk' portion of the B\&M simulation-- all particles beyond Neptune and interior to the Oort cloud ($a<3,000$~AU).

\cite{brasser2013} use a Nice model framework to populate the scattered disk and Oort cloud  and determine the relative sizes of these populations.
This Nice model migration includes a dynamical instability following the removal of the gas in the Solar System's protoplanetary disk, after the last encounter between the ice giants.
The planetary evolution track from \cite{nice} `Run A' was repeated, which starts with Neptune at 27.5~AU with an eccentricity of 0.3 and Uranus at 17.5~AU and an eccentricity of 0.2.
Uranus migrates outward to $\sim$19~AU and Neptune migrates to $\sim$31~AU over $\sim100$ Myr.
At the end of the simulation, Neptune is slightly beyond its true position, at 30.8 AU instead of 30.1 AU, however this was left unchanged during the simulation to avoid disrupting resonant objects and all objects were rescaled after completion.
The 30,000 test particle initial locations were $29<a<34$~AU (based on the final position of Neptune, this rescales to $28.2<a<33.2$~AU), $i<1^{\circ}$, and $e=0.15$.
The eccentricity of the test particle initial conditions results from the eccentricity of the outer planets \citep{nice}.
Each test particle was initially given unique position and velocity vectors.
For the scattered disk component, after the planet migration was completed, all test particles beyond 3,000 AU were removed from the simulation.
The particles and giant planets were integrated for an additional 3.8 Gyr with SWIFT RMVS3 \citep{swift}.
Because of significant scattering loss, after 1~Gyr and 3.5~Gyr, the remaining test particles were cloned three times to ensure a sufficiently well-sampled Kuiper belt (effectively 270,000 test particles).
The results from the end state of the 3.8 Gyr B\&M scattered disk simulations are utilized in this work; see \cite{brasser2013} for more details on the migration simulation.

\subsection{Additional Integrations}

In this work, the end state planet and test particle positions from the B\&M simulation were integrated forward 30~Myr in order to determine dynamical classifications.
Additional integrations were necessary because resonance classification requires more frequent output than the previous full simulation in order to conclusively classify the test particles.
The Sun, Jupiter, Saturn, Uranus, Neptune, and test particle end states from the previous simulation were provided as input for SWIFT RMVS4 \citep{swift}.
The particle positions were recorded every 300 years to ensure sufficient sampling of the resonant angle.
During the 30~Myr integration, the particles between $20<a<160$ AU were recorded, to focus on the scattering TNOs as well as some of the distant resonances.

Neptune's final semi-major axis from the B\&M simulations was $\sim$0.8 AU farther from the Sun than the true position of Neptune.
After the simulations completed, the semi-major axes of all objects in the B\&M simulation were adjusted to correspond with their locations if Neptune's semi-major axis were the current actual value, using a scale factor of 30.047/30.8 (a scaling of $<2.5$\%).
Because everything in the simulation is shifted by the same factor, the dynamics of all planets and test particles remain the same.
This adjustment was done to facilitate direct comparison with the Solar System, and for the remainder of this work, the positions discussed for Neptune, the test particles, and the mean-motion resonances in the simulations are the scale of the Solar System today.

\subsection{Particle Classifications}
\label{section_class}

\begin{figure}[h]
\begin{center}
\makebox[\textwidth][c]{\includegraphics[width=1.\textwidth]{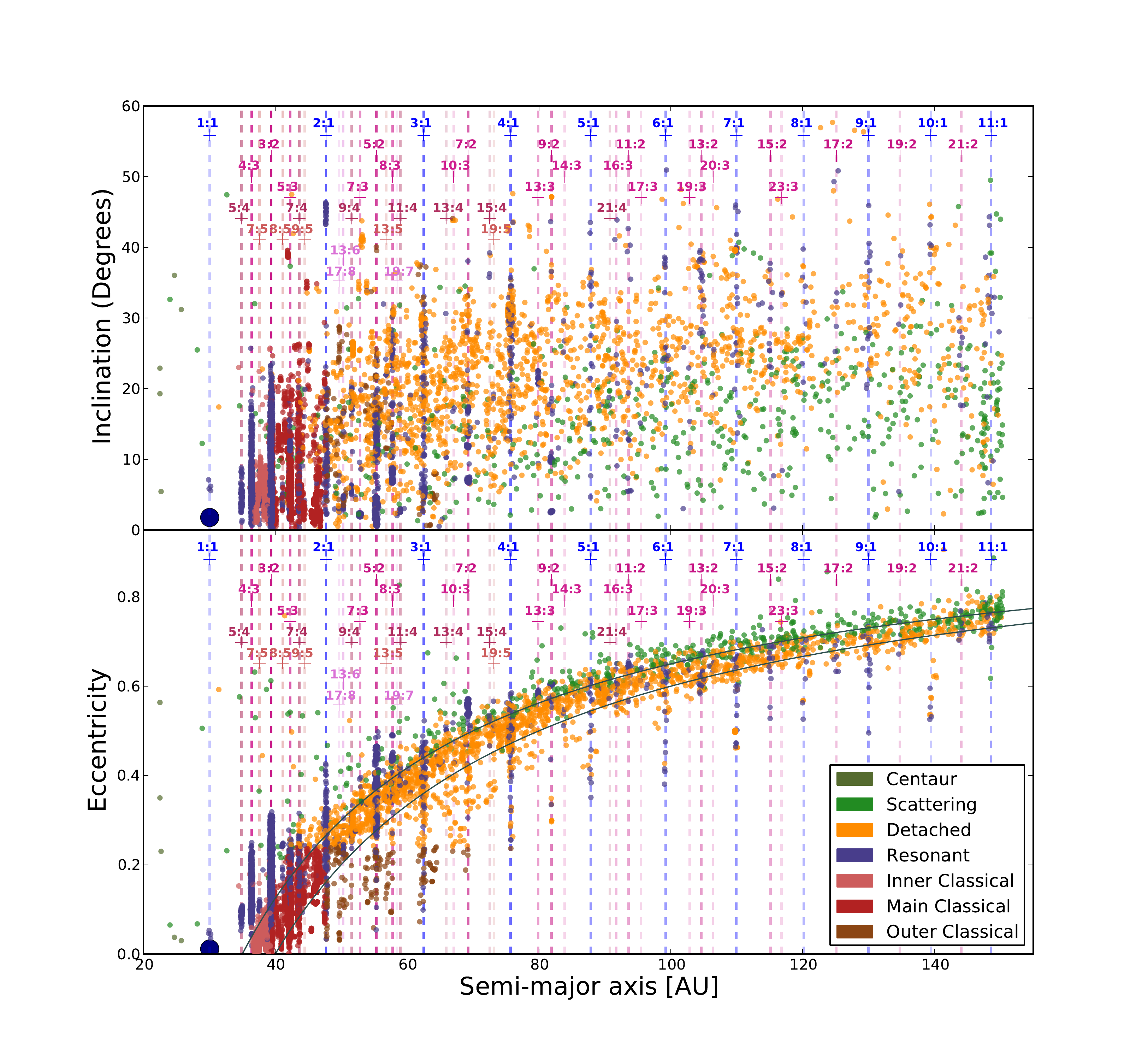}}
\caption{Test particle inclination, $i$, and eccentricity, $e$, distributions with semi-major axis, $a$, of the end state of the B\&M model.  The dashed lines mark resonances where more than two test particles are found.  The opacity of the dashed lines scales with the number of particles in the resonance.  The large number of inner and main classical objects is apparent.  The outer classical objects are consistent with emplacement through resonance dropout, similar to the slightly larger $e$ detached objects.  The solid lines indicate specific pericenter locations, $q$ of 35 and 40. Neptune is indicated by the large dark blue circle.}
\label{all_part}
\end{center}
\end{figure}

The subpopulations determined here are compared with survey detections from \cite{cfeps, hilat2016} and \cite{alexandersen}, so a classification scheme consistent with those survey classifications is used \citep[from][]{gladman2008}.
Figure \ref{all_part} shows a plot of all classified test particles.
The primary goal is to describe the behavior of the test particles at the start of the additional integrations, to characterize the B\&M model `end state.'

A classification of resonance requires an oscillation of the resonant angle, $\phi_{pq}$, over time, where $p$ and $q$ are integers, and $\phi_{pq}$ describes the $p$:$q$ mean-motion resonant angle.
Each particle whose semi-major axis is within 1.5 AU of a Neptune resonance location had the relevant resonant angle computed:

\begin{equation}
 \label{phi_equation}
 \phi_{pq} = p \lambda - q \lambda_N - (p-q)\varpi .
\end{equation} 

\noindent The particle mean longitude is $\lambda =  \Omega + \omega + \mathcal{M}$, $\Omega$ is the longitude of the ascending node, $\omega$ is the argument of pericenter, and the longitude of perihelion is $\varpi = \Omega + \omega$.
$\lambda_N$ refers to the mean longitude of Neptune.
If $\phi_{pq}$ oscillates instead of circulates, then the test particle is classified as resonant.

Diagnosing resonance based on a visual inspection is straightforward, but an automated detection method is required for large numbers of particles.
A spectrogram analysis was used on a series of windows to identify oscillation in $\phi_{pq}$ over time \citep{shankmanFFT}.
The behavior of $\phi_{pq}$ in overlapping windows of 5~Myr was analyzed using a fast Fourier transform (FFT).
If a particle's $\phi$ was resonant for all windows it was classified as stable; unstable particles only displayed oscillations in a subset of windows.
If an object was resonant, the libration amplitude (maximum to minimum $\phi_{pq}$ oscillation) and libration center (median value of $\phi_{pq}$) of the particle were calculated.

In this simulation all test particles beyond 34 AU, the original extent of the implanted disk, must have been scattered by Neptune.
`Scattering objects' specifically refers to the dynamically unstable objects in the simulation, particles that experience a change in $a \geq1.5$ AU in the first 10~Myr of the additional integration \citep{gladman2008}.
This is intended to refer to objects that are currently scattering, instead of the `scattered' or `scattered disk' classification which refers to objects that were scattered in the past, a difficult criterion to assess in real TNOs.
Objects that exhibit semi-major axis evolution and have $a < 30$ AU are classified as Centaurs.

If a particle is not currently scattering or resonant, then the particle is classified based on its current $a$, $e$, and $i$ values.
Objects are classified as main classical if they are between the 3:2 and 2:1 resonance, 39.4 AU -- 47.7 AU, and have eccentricities $e<0.24$.
Inner classical objects are found between Neptune and the 3:2 resonance, 30.04 AU -- 39.4 AU with eccentricities $e<0.24$.
Test particles beyond the 2:1 resonance at 47.7 AU in the same eccentricity range are classified as outer classical objects.
Detached objects are particles beyond Neptune with $e>0.24$ that are not scattering or resonant.

\clearpage{}
\vspace{50mm}

\section{Comparing the B\&M Model to the Solar System}
\label{comparing_section}

\subsection{Direct Model Comparison}

The subcomponents of the B\&M simulation results are directly compared to the CFEPS L7 model, the orbital element distributions and absolute population sizes of different TNO components based on the CFEPS detections and detection biases \citep{cfeps,gladman12} and other models based on CFEPS \citep{shankman13,pike2015,shankman2015}.
The L7 model is consistent with the CFEPS detections, but is not uniquely consistent; for example, an equally consistent model could contain additional populations at very large pericenters entirely undetectable by the surveys.
As a result, agreement between the B\&M simulation and the L7 model implies that the B\&M simulations must be consistent with the detections, while disagreement with the L7 model does not necessary require that the B\&M simulation results are inconsistent with real detections.
To mitigate this issue, the B\&M model is tested both against the L7 model and directly against the survey detections using the Anderson-Darling (AD) statistical test\footnote{See \cite{jones06} for an explanation of comparing cumulative parameters and how the AD statistic is used for rejection.}.

\subsection{Comparing the B\&M model to Real TNO Detections}
\label{survey_sim}

To facilitate direct comparison of the orbital distributions from the end state of B\&M to real TNO detections, an observational bias is applied to the B\&M simulation particles using a survey simulator.
The survey simulator uses survey detection characteristics (pointings, detection efficiencies, tracking efficiencies, etc.) to determine which input model objects would have been detected by the survey \citep{kavelaars2009}. 
The B\&M ``observed" test particles are compared to the detections from the CFEPS ecliptic \citep{cfeps} and high latitude \citep{hilat2016} surveys as well as the \cite{alexandersen} survey.
In both plots where B\&M simulation particles are compared to real TNOs, the TNOs are the characterized detections from these surveys.
To determine the acceptability of the B\&M particles as a model of the populations, the survey simulator-biased test particles are compared to the real TNO detections.

The B\&M simulation results were used as the input orbital model for the CFEPS survey simulator.
Conducting a survey simulator analysis of a simulation output requires a large number of orbit samples.
The simulation particles were cloned, preserving $a$, $e$, $i$, and in the case of resonant objects, the resonant angle ($\phi$), while randomizing the position angles ($\omega$, $\Omega$, and $\mathcal{M}$).
This cloning ensures a sufficient number of simulated detections; objects are randomly drawn until a larger sample than the known TNOs is `detected' to ensure the parameter space is sufficiently well sampled.

In order to determine the detectability of the B\&M orbital model, an absolute magnitude, $H$, distribution for the particles is assumed.
This is a proxy for object size, which is not directly measurable for unresolved objects.
$H$ distributions are typically parameterized as an exponential, equivalent to a single power law in diameter; in differential form the single power law has a single slope of $\alpha$:

\begin{equation}
\label{h_equation}
  dN/dH \propto 10^{\alpha H} .
\end{equation}

\noindent
\cite{shankman13,shankman2015} describes a broken $H$-distribution by joining two different differential single slopes, $\alpha_{\rm bright}$ and $\alpha_{\rm faint}$, at a specific magnitude, $H_{\rm transition}$.
In addition to the change of slope, they also propose a sudden drop (a `divot') in number density after the transition, parameterized as the contrast, $c$.
(A contrast of one has no drop and is referred to as a `knee' in the literature.)
The survey detections used in this work are primarily in $g$-band, so $H_g$, absolute magnitude in $g$ band is used.
The B\&M simulation particles are assigned an absolute brightness, $H_g$, from three different size distributions from the literature.
These are: a single slope of $\alpha=0.9$ as in \cite{gladman12}; a knee distribution with $\alpha_{\rm bright}=0.87$, $\alpha_{\rm faint}=0.2$, and $H_{g-\rm transition}=8.35$ as in \cite{fraser14}, converted to $g$ using $g-r$=0.65 from \cite{cfeps}; and a divot distribution with $\alpha_{\rm bright}=0.8$, $\alpha_{\rm faint}=0.5$, $H_{g-\rm transition}=9.0$, and a contrast $c=5.6$ as in \cite{shankman13}.
For the majority of the B\&M populations the choice of size distributions had no impact on the conclusions, so for these populations only the knee distribution is presented.

\begin{figure}[h!]
\begin{center}
\makebox[\textwidth][c]{\includegraphics[width=1.\textwidth]{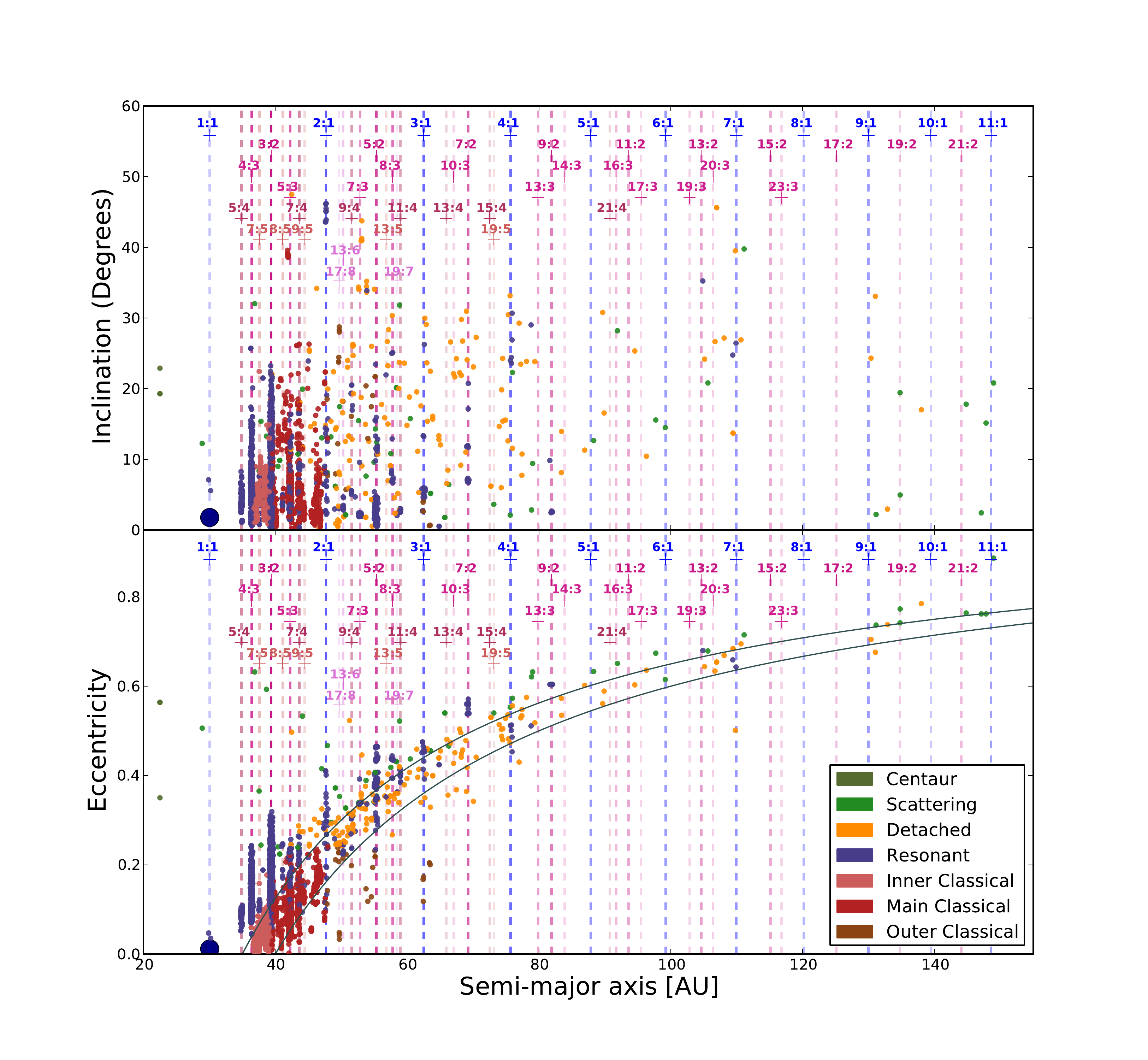}}
\caption{Similar to Figure \ref{all_part}, inclination, $i$, and eccentricity, $e$, distribution with semi-major axis, $a$, of the B\&M simulation end state biased using a survey simulator.  The 30,000 particles shown were `detected' by the survey simulator using the \cite{cfeps, hilat2016} and \cite{alexandersen} survey pointings with $H$ magnitudes randomly assigned from a single slope $H$-magnitude distribution with $\alpha=0.9$.  This plot includes only detections with $H_g<8$, which roughly corresponds to $>170$ km in diameter.  The significant selection effects of TNO surveys are apparent; the inner classical and close resonances are much easier to detect compared to more distant populations.  The knee and divot distributions show qualitatively similar detection biases.}
\label{survey_bias}
\end{center}
\end{figure}

The survey simulator biased B\&M model reflects the detectability of different populations, see Figure \ref{survey_bias}.
The selection biases which complicate TNO population studies are apparent when comparing Figure \ref{all_part} and Figure \ref{survey_bias}. 
The closest objects dominate the simulated detections.
The fraction of detached objects detected is significantly smaller than other populations, because of their large pericenters.
The choice of size distribution model affects the expected number of detections roughly as a function of perihelion distance.
The single slope produces more small object detections per each large object, so for the same number of simulated detections, a single slope distribution results in more low-$a$ and large-$H$ detections, while a knee or divot distribution is more likely to have larger-$a$ and low-$H$ detections.
The survey simulator biased results for several TNO subpopulations are presented in Section \ref{classifying}.

\clearpage{}
\vspace{50mm}

\section{Results: Populations of the Outer Solar System}
\label{classifying}

Using the methods described in Section  \ref{section_class}, the end-state orbits of the test particles in the B\&M simulations were placed into orbital classes.  
The full population statistics for the B\&M model are summarized in Table \ref{classifications} and the model objects' orbital distribution is plotted in Figure \ref{all_part}.
Several subpopulations are discussed in detail here.

\begin{table}[h]
\setlength{\tabcolsep}{2.5pt}
\caption{Test Particle Classifications}
\label{classifications}
\begin{center}
\begin{tabular}{ l c  c  c  c  c  c  c c  c c c c }
\hline\hline
Classification & Number of Particles & Fraction of Total   \\\hline
Resonant  & 3,910 &   42\% \\
Inner Classical & 871 & 9\% \\
Main Classical & 1,943 & 21\% \\
Outer Classical & 181 &  2\% \\
Scattering  & 535 &  6\%  \\
Detached  & 1,921 &   21\% \\
Centaurs & 5 & 0.05\% \\ \hline 
Total Particles  & 9,366 & 100\%   \\
\hline
\end{tabular}
\end{center}
\end{table}

\subsection{Scattering Objects}
\label{scatter_section}

\begin{figure}[h!]
\begin{center}
\includegraphics[width=.5\textwidth]{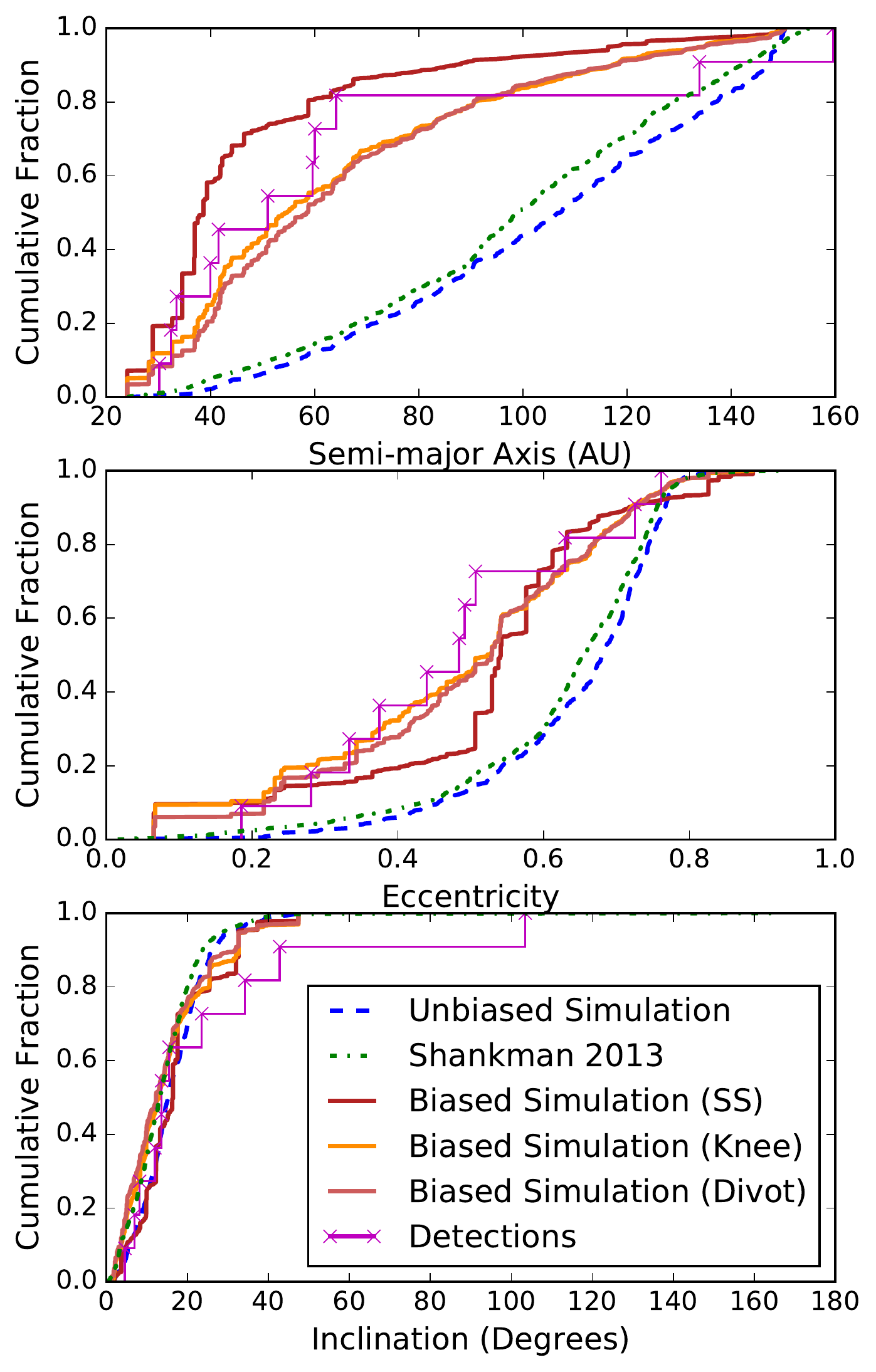}
\caption{The cumulative fraction of scattering objects in the B\&M simulation with the $a$, $e$, and $i$ values are indicated with the blue (dashed) line.  The green (dash-dot) line indicates the model used to represent the unbiased model distribution from \citet{shankman13,shankman2015}, which is consistent with the blue (dashed) B\&M simulation end state.  The B\&M particles were assigned three different $H$-magnitude distributions: the single slope (SS), knee, and divot.  The particles were then biased using the survey simulator.  The magenta `x' marks indicate actual detections from the surveys simulated, for comparison with the B\&M biased simulation results.  The observed orbital element distributions are better matched by the knee or divot size distributions than the single slope.}
\label{scatter}
\end{center}
\end{figure}

The B\&M scattering objects populate the considered region from 24--155 AU at a nearly constant rate over semi-major axis.
These objects are shown in Figure \ref{scatter}, and their distribution in $a$, $e$, and $i$ is statistically consistent with the model of scattering objects used by \cite{shankman13,shankman2015}.
The \cite{shankman13,shankman2015} scattering object model used orbital parameters based on the simulation results from \cite{kaib11}.
The scattering objects from \cite{kaib11} were produced by a significantly different planetary migration and evolution scenario, however the signatures in the orbital structure produced by the specific migration in the scattering objects are not statistically distinguishable.
Based on comparing these scattering object orbital element distributions, we conclude, as in \cite{shankman13}, that the specifics of planetary migration do not strongly affect the scattering population.

Of the populations considered in this work, the scattering objects are the most sensitive to the choice of size distribution.
The detectability of scattering objects with the divot, knee, and single slope size distributions applied to the B\&M simulation scattering objects is shown in Figure \ref{scatter} (see Equation \ref{h_equation} for details).
Previous work from \cite{shankman13,shankman2015} finds that the single slope is rejectable for the scattering objects, and the divot is the preferred model (although several knee distributions are acceptable).

The biased B\&M simulation with the single slope, knee, and divot $H$-distributions all provide a statistically acceptable semi-major axis distribution match for the real detections.
The single slope results in a significantly worse eccentricity distribution, nearly rejectable by the AD statistic.
The knee and divot $e$-distributions are non-rejectable.
The B\&M model does not include any retrograde objects; these retrograde objects are returning Oort cloud objects \citep{brasser12} which were excluded from the simulation, so the lack of retrograde objects does not invalidate the model.
When the detected retrograde TNO (from \cite{hilat2016} at $i>90^{\circ}$ in Figure \ref{scatter}) is excluded from the analysis, the inclination distribution of the three $H$-distributions all provide an acceptable match.
Based on the B\&M model of the $a$, $e$, and $i$ distribution of the scattering objects, a knee \citep{fraser14} or divot \citep{shankman13} size distribution provides a significantly better representation of the real detections than a single slope $H$-distribution.
The orbital parameters of the B\&M scattering objects are consistent with the parameters from \cite{kaib11}, confirming that properties of the scattering population are not particularly dependent on the specifics of the scattering event \citep{shankman13}.
The scattering objects in the B\&M simulation with a knee or divot size distribution provide a good model of this population.

\subsection{Resonant Test Particles}

The B\&M simulation contains 42\% resonant test particles.
The B\&M model inner resonances are overpopulated relative to the Solar System.
This analysis of B\&M resonant test particles focuses on resonances beyond the main classical belt, including the 2:1 resonance.

The orbital distribution of the B\&M resonant particles at smaller semi-major axes, such as the 3:2, 4:3, and 5:4, are unlikely to represent the real object distributions because of the initial B\&M simulation design.
This is likely a result of the extended disk of particles before scattering, which results in some unrealistic sweeping capture.
When the test particles with initial semi-major axes interior and exterior to the final position of Neptune are considered separately, the $a>30$~AU particles are captured into these closer resonances twice as efficiently as the $a<30$~AU particles. 
However, the capture efficiency beyond the 2:1 resonance does not depend on initial particle $a$, so all particles are useful in analyzing the distant populations.
Many of the more distant resonant populations in the B\&M simulation provide an excellent match to observations, as these resonances are populated by captured scattering objects which are well represented by the B\&M model, in part due to the generic nature of scattering (see Section \ref{scatter_section}).

The resonance occupation at the end state of the B\&M simulations is presented in Table \ref{reslist}.
The resonances with population estimates from the comparison surveys are included here; resonances populated in the B\&M simulation but not identified in the surveys are excluded.
The number of test particles in each resonance is reported, as well as the fraction classified as `stable'; this requires that the test particle be resonant for the entire classification integration of 30 Myr.
Unstable resonant particles are resonant for a minimum of 5 Myr.
To assess the B\&M simulation, these resonant B\&M populations are compared to debiased survey results from the literature, as well as biased using a survey simulator for comparison with real resonant object detections.

\begin{table}[h]
\setlength{\tabcolsep}{2.5pt}
\caption{Resonance Occupation}
\label{reslist}
\begin{center}
\begin{tabular}{ l c  c  c  c  c  c  c c  c c c c }
\hline\hline
Resonance & Semi-Major Axis & Number of & Fraction \\ 
$p$:$q$ & (AU) & Test Particles &  Stable$^a$ \\ 
\hline
1:1 & 30.05 & 4 & 100\%  \\ 
5:4 & 34.87 & 64 & 100\%  \\ 
4:3 & 36.40 & 588 & 100\%  \\ 
 3:2 & 39.37 & 1640 & 99\%  \\ 
5:3 & 42.24 & 217 & 100\%  \\ 
7:4 & 43.63 & 67 & 100\%  \\ 
2:1 & 47.70 & 111 & 97\%  \\ 
7:3 & 52.86 & 39 & 100\%  \\ 
5:2 & 55.35 & 337 & 99\%  \\ 
3:1 & 62.50 & 77 & 85\%  \\ 
4:1 & 75.71 & 70 & 80\%  \\ 
5:1 & 87.86 & 19 & 84\%  \\  
\hline
\end{tabular}
\end{center}
\footnotesize{$^a$Particles are considered stable when resonant for the 30 Myr integration.}\\
\end{table}

\subsection{Comparing Population Sizes: Resonant and Scattering Objects}
\label{size_section}

A successful model of planetary migration should reproduce the relative population sizes of the Kuiper belt resonances.
The B\&M simulation population ratios are presented in two ways in this section.
In Table \ref{survey_res}, the end state B\&M populations are compared to published debiased literature population estimates.
The `L-Scale Factor', literature scale factor, in this table shows the factor the simulation would need to be increased by in order to match the literature estimate, and the 95\% uncertainty in the literature population estimate is translated into the scale factor 95\% uncertainty.
For Table \ref{survey_res_biased}, the B\&M model is biased using the survey simulator and compared to the number of real TNO detections from the surveys.
The `B-Scale Factor', or biased scale factor, is the median number of times the model of that dynamical sub-class must be sampled in order to produce the number of real detections listed, with 95\% confidence intervals from the population estimate distributions.
These scale factors indicate the factor by which the starting disk would have to be increased in order to result in the appropriate population size.
In each table, when the populations have similar scale factors (with overlapping uncertainty) this indicates that these populations are produced consistently within the B\&M model and scale factor agreement between the tables indicates that the population size is consistent by both methods.

The population sizes of the B\&M model sub-components are compared to population estimates of the resonances from carefully characterized surveys \citep{cfeps, alexandersen, bannister2016, hilat2016}.
The population estimates \citep{gladman12,pike2015,alexandersen,volk2016} were calculated by creating a parametric model of the resonance, then using the survey simulator to forward bias the model to statistically compare it to the observations.
(See \cite{gladman12} or \cite{volk2016} for a detailed explanation of the paramaterization.)
The size of the underlying population necessary to generate the number of detections in the survey is the estimated size of the population.
The population estimates for many resonances explored by these surveys are summarized in Table \ref{survey_res}.

\begin{table}[h!]
\renewcommand{\thetable}{\arabic{table}}
\setlength{\tabcolsep}{2.5pt}
\renewcommand{\arraystretch}{1.1}
\caption{Literature Estimates of Resonant Populations from Surveys}
\label{survey_res}
\begin{center}
\begin{tabular}{ l c  c  c  c  c  c  c c  c c c c }
\hline\hline
Resonance & $a$ & B\&M Simulation & Population Estimate & Survey; Source \\
$p$:$q$ & (AU) & L-Scale Factor$^a$  & N($H_g<8$) &   \\\hline

1:1 & 30.05 &  5.0$^{+20}_{-4}$ & 10$^{+40}_{-9}$  & \cite{alexandersen}$^b$ \\ \arrayrulecolor{gray}\hline

5:4 &  34.87 & 0.4$^{+2.4}_{-0.36}$ & $10^{+60}_{-9}$  & CFEPS; \cite{gladman12} \\ \hline

4:3 & 36.40 &  0.4$^{+0.4}_{-0.32}$  & $70^{+100}_{-50}$  &  CFEPS; \cite{gladman12}  \\ \hline

\multirow{ 3}{*}{3:2} & \multirow{ 3}{*}{39.37} &   1.6$^{+0.6}_{-0.6}$  & $1200^{+500}_{-400}$    & CFEPS; \cite{gladman12} \\
     & &1.4$^{+0.6}_{-0.4}$ & $1100^{+400}_{-300}$   &  \cite{alexandersen} \\
     & &1.2$^{+0.4}_{-0.4}$ & $900^{+330}_{-270}$   & CFEPS+OSSOS$^c$; \cite{volk2016} \\ \hline
     
5:3 & 42.24 & 12$^{+12}_{-6}$  & $450^{+470}_{-280}$    & CFEPS; \cite{gladman12}  \\ \hline

7:4 & 43.63 &  40$^{+60}_{-20}$  & $300^{+400}_{-200}$  & CFEPS; \cite{gladman12}  \\ \hline

\multirow{ 2}{*}{2:1} & \multirow{ 2}{*}{47.70} & 20$^{+10}_{-10}$  & $340^{+200}_{-220}$  &  CFEPS; \cite{gladman12}  \\
     & &20$^{+10}_{-10}$ & $360^{+230}_{-180}$   & CFEPS+OSSOS$^c$; \cite{volk2016} \\ \hline
     
7:3 & 52.86 & 20$^{+40}_{-10}$ & $320^{+760}_{-270}$  & CFEPS; \cite{gladman12}  \\\hline

\multirow{ 2}{*}{5:2} & \multirow{ 2}{*}{55.35} &  6$^{+8}_{-4}$ & $1100^{+1400}_{-700}$   &  CFEPS; \cite{gladman12}  \\
     & & 4$^{+4}_{-2}$ & $770^{+680}_{-420}$   & CFEPS+OSSOS$^c$; \cite{volk2016}\\ \hline
     
\multirow{ 2}{*}{3:1} & \multirow{ 2}{*}{62.50} &  10$^{+20}_{-8}$  &  $340^{+800}_{-290}$  & CFEPS; \cite{gladman12}  \\
   & & 6$^{+8}_{-4}$  & $220^{+270}_{-150}$  &  \cite{alexandersen} \\ \hline
   
4:1 & 75.71 & 3$^{+10}_{-3}$ & $80^{+360}_{-80}$   &   \cite{alexandersen} \\ \hline

5:1& 87.89 & 240$^{+420}_{-180}$ & 1900$^{+3300}_{-1400}$   & CFEPS; \cite{pike2015}   \\
\arrayrulecolor{black}\hline
\end{tabular}
\end{center}
\footnotesize{Note that the 5:3, 4:3, and 3:2 resonances are overpopulated in the B\&M simulation.}\\
\footnotesize{$^a$The `L-Scale Factor' (literature scale factor) indicates how much the B\&M simulation (with $H_g<8.35$) must be scaled up to match the population estimates (Population Estimate scaled to 8.35 $\div$ Number of B\&M model objects).  The uncertainties are the propagated 95\% uncertainties from the population estimates.}\\
\footnotesize{$^b$Population estimate is for the stable Neptune Trojans, as all of the B\&M test particles in this resonance are stable.}
\footnotesize{$^c$OSSOS: the Outer Solar System Origins Survey \citep{bannister2016}}
\end{table}

\vspace{50mm}

As discussed in the previous section, the B\&M simulation was highly efficient at populating resonances interior to the classical Kuiper belt compared to those beyond the 3:2, causing an overpopulation in the 5:4, 4:3, and 3:2 resonances (see Table \ref{survey_res}).
These low-$a$ resonances include a component captured by resonance sweeping.
The initial planetesimal disk extended to 33.2 AU, contrary to expectations about the real proto-planetesimal disk.
The starting conditions for Neptune in the B\&M simulation placed the 5:4 resonance and the 4:3 resonance within the initial disk before planetary migration.
The 3:2 resonance was just beyond the original extent of the implanted disk, but the test particles had sufficient eccentricity to reach an apocenter crossing this resonance, so sweeping may still be effective.
The B\&M 5:4, 4:3, and 3:2 resonances are the only resonances that include a significant number of particles swept into resonances, resulting in too high a capture efficiency, so these populations are not representative of the real TNOs.

\clearpage{}

The number of objects in many B\&M resonant populations (2:1, 7:3, 5:2, 3:1, 4:1) and the population estimates from \cite{gladman12}, \cite{volk2016}, and \cite{alexandersen} are consistent.
The L-scale factors for these populations in Table \ref{survey_res} are consistent with each other and notably agree with the observational survey results from \cite{gladman12} and \cite{volk2016} that the 5:2 resonance has a large population.

\setcounter{table}{4}
\begin{table}[h!]
\setlength{\tabcolsep}{2.5pt}
\addtocounter{table}{-1}
\caption{Populations from the B\&M model biased using a survey simulator.  The knee $H$-distribution is presented because the effects of different $H$-distributions are minimal.}
\label{survey_res_biased}
\begin{center}
\setlength{\tabcolsep}{0.5cm}
\begin{tabular}{ p{1.4cm} p{1.cm}p{1.5cm}p{1.cm}}
\hline\hline
Sub- & $a$ & B\&M/Survey & Survey  \\
population ($p$:$q$) & (AU) &  B-Scale Factor$^a$& Detections$^b$  \\\hline
1:1 & 30.05 & 4 $^{+ 20 }_{- 4 }$  & 1 \\
5:4 & 34.87 & 0.2 $^{+ 0.8 }_{- 0.2 }$  & 1 \\
4:3 & 36.40 & 0.2 $^{+ 0.3 }_{- 0.1 }$   & 6 \\
3:2 & 39.37  & 0.8 $^{+ 0.3 }_{- 0.2 }$  & 42 \\
5:3 & 42.24 & 2.5 $^{+ 1.8 }_{- 1.2 }$ & 12 \\
7:4 & 43.63 & 5 $^{+ 6 }_{- 3 }$  & 5 \\
2:1 & 47.70 & 10 $^{+ 8 }_{- 5 }$   & 9 \\
7:3 & 52.86 & 3 $^{+ 6 }_{- 2 }$  & 2 \\
5:2 & 55.35 & 1.4 $^{+ 1.4 }_{- 0.7 }$  & 8 \\
3:1 & 62.50 &  11 $^{+ 15 }_{- 7 }$  & 4 \\
4:1 & 75.71 & 6.6 $^{+ 31 }_{- 6.4 }$  & 1 \\
5:1 & 87.89 & 180 $^{+ 330 }_{- 140 }$  & 3 \\
Scattering & 20--155  & 5 $^{+ 4 }_{- 2 }$   & 12 \\
\hline
\end{tabular}
\end{center}
\footnotesize{$^a$The `B-Scale Factor' (biased scale factor) is the number of times the B\&M model of the selected population with the knee size distribution must be sampled by the survey simulator to generate the number of detections found by the surveys.  The 95\% uncertainty quoted is calculated by randomly resampling the population.  The simulated detections were counted to $H_g<$8.35, the knee in the size distribution.}\\
\footnotesize{$^b$Total number of detections as found by \cite{gladman12}, \cite{hilat2016}, and \cite{alexandersen} surveys}.\\
\end{table}

To avoid relying on the specific orbital distributions from the survey population models, the B\&M model is biased using the survey simulator (see Section \ref{survey_sim}).
Some of the published survey population estimates have large 2$\sigma$ uncertainties.
The biased B\&M model output is presented in Table \ref{survey_res_biased}.
The scale factors for the knee $H$-magnitude distribution are presented; the choice of size distribution made no statistically significant difference in the results.

The $n$:1 resonances are likely populated by the capture of scattering objects, so if the B\&M model accurately reproduces the scattering population it should reproduce the $n$:1 population statistics as well.
The relative sizes of the scattering object population and some of the $n$:1 resonance populations  in Table \ref{survey_res_biased} are consistent with observations.
For the real survey detections, the scattering / 3:1 / 4:1 populations have a ratio of 12 / 4 / 1; the biased B\&M simulation gives a consistent ratio of 20 / 5 / 1 before conversion to the scale factor.
The B- and L-scale factors ($\sim$5-10) between the detections and the model are within uncertainties for these populations.
The B\&M model for the scattering objects, 3:1, and 4:1 resonance is self-consistent.

\subsection{The Large Population of 5:1 Resonators}
\label{five_one}

\cite{pike2015} investigated the 5:1 Neptune resonance using both real and simulated detections.
The three real TNO detections were found by CFEPS \citep{cfeps, hilat2016} and a population model was created based on the constraints provided by the detections and the survey characterization.
The parametric model of the 5:1 resonance from \cite{pike2015} is consistent with a minimum population estimate based on the orbital distribution of 5:1 resonators from the B\&M model.

In the B\&M evolution, the 5:1 resonance is populated by the same mechanisms as the 3:1 and 4:1 resonances, but both the L- and B-scale factor comparisons show that the B\&M model significantly under-predicts this population compared to survey estimates and discoveries \citep{pike2015}.
If the B\&M model is correct, the efficiency of detection for the 5:1 resonance would be extremely low; producing three 5:1 detections in the surveys would result in $\sim1,000$ scattering object detections.
In order to be consistent with the B\&M scattering / 3:1 / 4:1 populations, the 5:1 B\&M population would need to be $\sim20-100\times$ larger, requiring an unreasonably large starting planetesimal disk.
The detections in the 5:1, and thus the large population in that resonance, requires a different population source.
This confirms that the extremely large population estimate for the 5:1 found by \cite{pike2015} is unexplained by the currently explored models.

The 5:1 resonance is significantly underpopulated in the B\&M simulation compared to expectations from survey results.
\cite{pike2015} predict an enhancement of a factor of $\sim50-100$ compared to the local scattering objects, but this is not confirmed in typical Kuiper belt formation models \citep{hahnmalhotra05,nice}.
The non-resonant populations in the B\&M model within $\pm5$ AU of the 5:1 resonance (89 AU) include 54 scattering and 159 detached test particles. 
This is a linear density of 5 scattering and 16 detached objects per AU.
The B\&M 5:1 resonance has 16 particles, with an approximate width of 1~AU it is $\sim3$ times denser than the scattering object population and comparable in density to the detached population, inconsistent with the enhancement predicted by \citep{pike2015}.

 \begin{figure}[h!]
\begin{center}
\includegraphics[width=.52\textwidth]{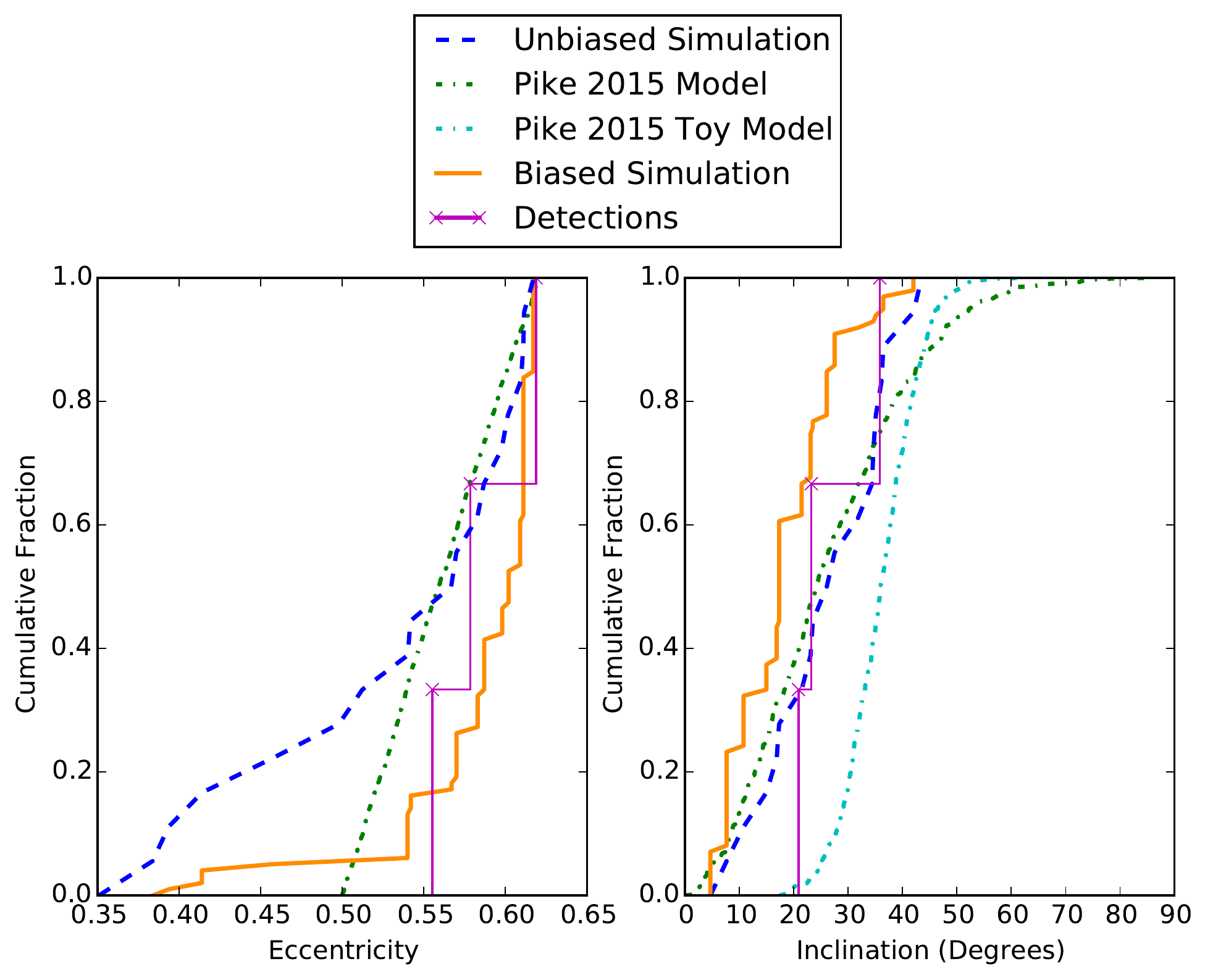}
\caption{These are the 5:1 resonators in the B\&M simulation (blue dashed) as well as the parametric model from \citet[green dash-dot]{pike2015} and the toy inclination model from \citet[turquoise dash-dot]{pike2015}.   The toy model was proposed to explore the possibility of an exclusively large-$i$ population, $\sigma_i=7^{\circ}$ and $\mu=35^{\circ}$. The magenta `x' detections are the real 5:1 objects discovered in \cite{cfeps, hilat2016}.  
The model eccentricity distribution has an appropriate upper limit, but the B\&M results suggest the eccentricity is not truncated at 0.5.  
Also plotted are the results of biasing the B\&M simulation using the survey simulator and the knee $H$-magnitude distribution.  
The biased population does not include a significant low-$e$ component, reflecting the difficulty of observing the low-$e$ particles at large-$a$.
The preferred inclination distribution with $\sigma_i=22^{\circ}$ is an acceptable match for the B\&M test particles.}
\label{fivetinos}
\end{center}
\end{figure}

The eccentricity and inclination distributions of the 5:1 resonators from the B\&M model and the parametric model distributions from \cite{pike2015} are similar, shown in Figure \ref{fivetinos}.
The upper limit of the eccentricity distribution is a good match, however the B\&M simulation has test particles in the range $0.35<e<0.62$, extending to lower eccentricities than the model from \cite{pike2015}.
If the B\&M simulated population distribution is representative of the real 5:1 resonators, the \cite{pike2015} eccentricity distribution underestimates the population size by $\sim$35\%.
The AD statistic shows that the eccentricity distribution of the biased B\&M 5:1 resonators is consistent with detections.
Inclination distributions have been found to be well modeled using the distribution $\sin(i)\times \exp\left(\dfrac{-(i-\mu)^2}{2\sigma_i^2}\right)$, including a peak shift, $\mu$, as well as the width, $\sigma_i$ \citep{brown2001,gulbis,pike2015}.
The preferred inclination width $\sigma_i=22^{\circ}$ from \cite{pike2015} is consistent with the B\&M model results.
The toy model from \cite{pike2015} with an inclination width of $\sigma_i=7^{\circ}$ shifted to a center of $\mu=35^{\circ}$ under-predicts the $i<35^{\circ}$ portion of the $i$-distribution in the B\&M model.
Overall, the $e$- and $i$- distributions of the B\&M test particles in the 5:1 resonance compares favorably to the preferred parameterization of the 5:1 resonators by \cite{pike2015}.

The three 5:1 detections in CFEPS imply a large population that is inconsistent with the B\&M model; the source of the 3$\sigma$ discrepancy between these published results is unclear and will only be resolved through additional observations or modeling.
Minor model differences, such as the assumptions by \cite{pike2015} of a smaller asymmetric libration fraction and eccentricity distribution than the B\&M model, only result in a larger population estimate, and therefore cannot reconcile the two estimates.
It is possible that some additional dynamical population mechanism not included in B\&M emplaces objects into the 5:1 resonance.
The discovery of new 5:1 resonators (or a lack of new resonators) in ongoing and future surveys, such as OSSOS \citep{bannister2016}, may identify other issues with the 5:1 resonator model parameterization or reduce the population estimate.
The large population in the 5:1 resonance remains unresolved, with the B\&M simulation results unable to explain this enhancement identified from survey results.

\subsection{Initial Disk Mass}
\label{mass_section}

For the \cite{nice} simulation, planetary migration is driven by an initial disk estimated to be $\sim35$~M$_\oplus$.
Nice model simulations have used a range of masses, $\sim18-35$~M$_\oplus$ \citep[e.g.][]{nesvorny_morbidelli2012}.
The slow evolution of the giant planets is driven by a gradual leaking of planetesimals until the instability caused by giant planet encounters begins.
These encounters cause significant mass loss in the disk; a factor of $\sim$10 removal of the disk mass during these giant planet encounters is consistent with the rapid decay of the mass in a Nice model scenario \citep{booth2009}.
The B\&M simulation begins after the last encounter between the ice giants, after significant mass loss from the initial $\sim35$~M$_\oplus$ disk has occurred.

 \begin{figure}[h!]
\begin{center}
\includegraphics[width=.52\textwidth]{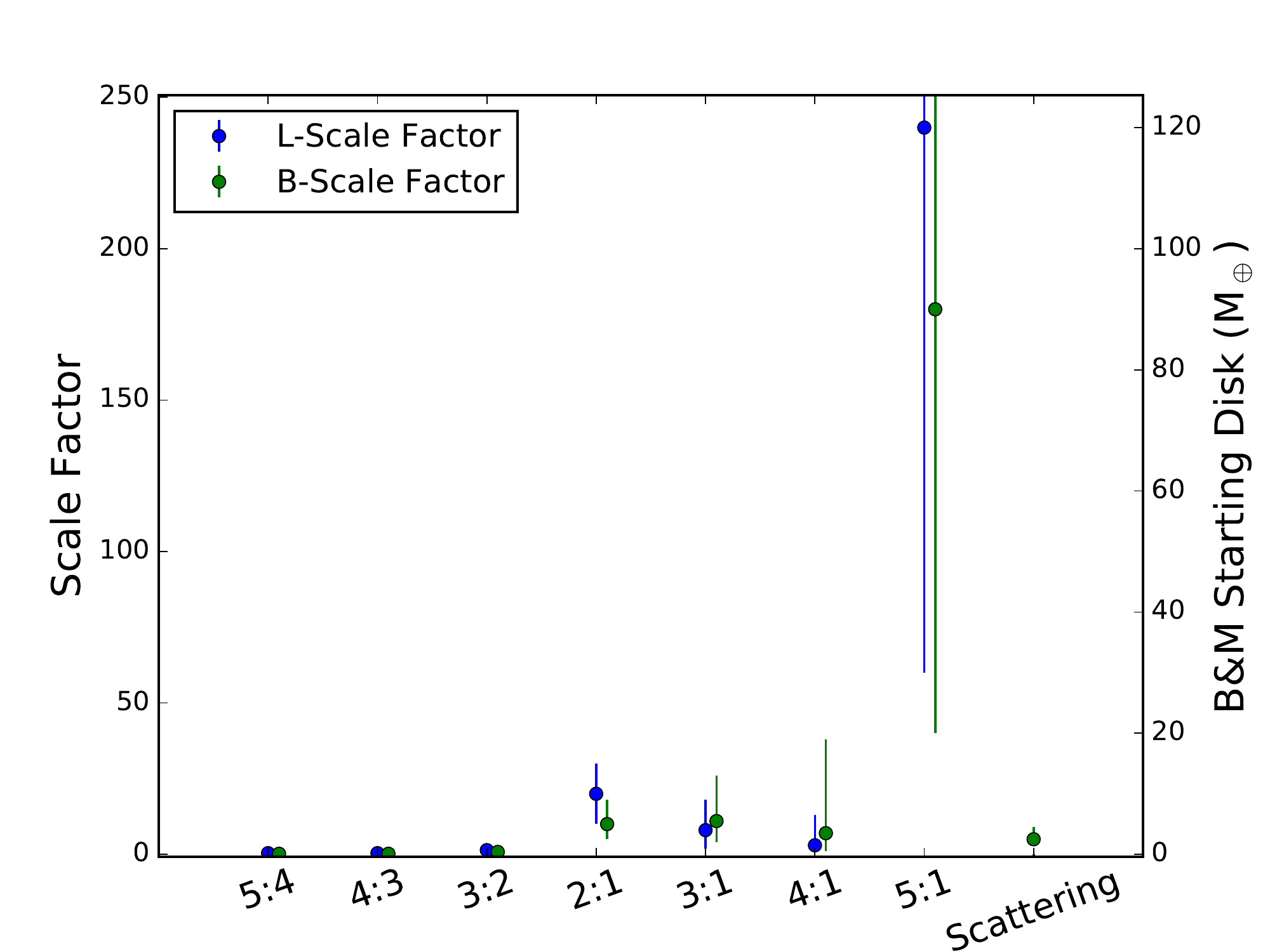}
\caption{The B- and L-scale factors and the corresponding starting disk masses are shown here (from Tables \ref{survey_res} and \ref{survey_res_biased}).  As in Section \ref{size_section}, the L-scale factor compares the size of the B\&M model to the size of literature models, and the B-scale factor compares the survey simulator biased B\&M model to real TNO detections from surveys.  The disk mass is calculated based on the scaling of the starting disk particles in order to emplace a sufficient number of TNOs in the particular population.  The 5:4, 4:3, and 3:2 resonances all have too small a scale factor and starting disk mass, while the 5:1 resonance starting disk mass is much too large.  The 2:1, 3:1, 4:1, and scattering populations predict starting disk masses that are consistent with expectations.}
\label{mass}
\end{center}
\end{figure}

With the assumption of a size distribution for the particles in the B\&M model (see Section \ref{survey_sim}), the mass of the planetesimal disk at the start of the B\&M simulation can be determined based on this distribution and the number of particles.
This mass may need to be scaled up by some factor in order to correctly reproduce the final mass in the Kuiper belt.
The scale factors have already been calculated in Section \ref{size_section}.
For planetesimals with a knee $H$-distribution, because of the shallow slope after $H_g$=8.35 (the location of the knee), the objects larger than the knee effectively contain all of the mass.
Assuming an albedo of 5\%, $H_g$=8.35 corresponds to an object diameter of 160~km.
For 30,000 particles larger than the knee in the size distribution and up to 2400~km diameter ($\sim$Pluto), assuming a density of 1.5 g/cm$^3$, the mass of the disk of particles is 3.1$\times10^{23}$~kg.
Because of the cloning during the simulation, the starting planetesimal disk is increased by a factor of 9 (to 270,000 particles), so the mass is 3$\times10^{24}$~kg, $\sim$0.5~M$_\oplus$.
This planetesimal mass can be scaled to match the number of starting particles needed to produce the survey detections; for the well-modeled populations this is a factor of 5-10 (see Section \ref{size_section}), which corresponds to a starting disk mass of 2-5~M$_\oplus$ for the B\&M simulation.
Figure \ref{mass} shows the range of scale factors and masses for the different populations.
This post-instability mass of 2-5~M$_\oplus$ predicted by the B\&M simulation results from depletion of the initial planetesimal disk; in Nice model type scenarios this initial mass is 18-35~M$_\oplus$ implying a factor of $\sim10$ depletion during the instability, consistent with expectations based on \cite{booth2009}.
The initial disk masses of $\sim20$~M$_\oplus$ from \cite{nesvorny_morbidelli2012} resulting from wide initial configurations of the planets requires less depletion, but only a factor of $\sim$4--10.
Therefore we find that the mass scale required for Nice model type scenarios is consistent with the observed Kuiper belt today.

\vspace{2cm}

\section{Discussion and Conclusions}
\label{conclusions}

In this work, the results of the B\&M Nice model simulation were compared with well-characterized surveys: CFEPS \citep{cfeps, hilat2016}, OSSOS \citep{bannister2016}, and \cite{alexandersen}.
These comparisons test the effectiveness of a cosmogonic model involving Neptune scattering (similar to the Nice model) in creating a detailed Kuiper Belt that matches the reality observed today.
For some resonant populations, this Nice model scenario provides a reasonable match, particularly resonances populated through scattering capture (3:1, 4:1), but for others the model does not produce an acceptable population distribution (3:2, 4:3, 5:4).

The closest resonances (3:2, 4:3, and 5:4) are populated too efficiently in the B\&M simulation.
This results from the extended intial planetesimal disk, with particles from 28.2 to 33.2 AU.
The particles with initial conditions exterior to Neptune's final semi-major axis were captured into these resonances with twice the efficiency of particles which started at smaller semi-major axes.
Previous work has argued that the planetesimal disk could not have extended beyond $\sim30$~AU in order to halt Neptune's migration at its current position \citep{gomes2004}.
However, this work shows that the planetesimal disk also could not have extended beyond $\sim30$~AU with constant density because capture into the 3:2, 4:3, and 5:4 resonances would have been too efficient compared to the populations observed in these resonances today.

The size of the B\&M population in the 5:1 resonance is severely underpopulated compared to the other resonant populations and the scattering objects in the B\&M model.
\cite{pike2015} calculated that the 5:1 population is likely as populous as the 3:2 resonance based on TNO detections in \cite{cfeps, hilat2016}.
The B\&M simulation produces acceptable population ratios between the 3:1, 4:1, and scattering populations.
The 5:1 resonance is populated in the same manner as the 3:1 and 4:1 resonances, however the B\&M simulation produces far too few 5:1 resonators compared to the number of detected TNOs from CFEPS and the model by \cite{pike2015}.
The B\&M simulation results were compared with the de-biased CFEPS models of the 5:1 and scattering population, and the B\&M simulation results were biased using the CFEPS survey simulator and compared to the detections.
In both comparisons, the small size of the 5:1 resonance in the B\&M simulation is incompatible with the three 5:1 detections.
The inclusion of an additional planet in the Kuiper belt region would only result in smaller populations in the 5:1 resonance \citep{lykawka2008}.
In order to resolve this discrepancy between published models, the population in the 5:1 resonance must have an additional source and population mechanism or additional survey results may reveal additional information about the population which reduces the detection frequency and population estimate.
The large observed population of 5:1 resonators in the Kuiper belt remains unexplained.

The resonant objects are a good test of the accuracy with which the B\&M simulation reproduces the Kuiper belt; they account for 42\% of the test particles and are sensitive to the specifics of migration \citep[e.g.][]{gomes2004,hahnmalhotra05,nesvorny2015b,nesvorny2015a}.
Several of the resonant test particle populations do not have a hot enough inclination distribution, a common issue in dynamical evolution models \citep{gladman12}, however the more distant resonance inclination distributions are well matched by the B\&M simulation.
The simulated eccentricity distributions are consistent with the data, and may provide a useful model basis for the more distant resonances which have significantly fewer detections.
These high-$a$ resonances in the B\&M simulation often lack a low-$e$ component, typically omitted from population models for detectability concerns \citep{gladman12,pike2015,volk2016,alexandersen}, which suggests that the practice of omitting these nearly undetectable low-$e$ objects from population models does not invalidate these population estimates.

The B\&M simulation results in an acceptable model of the populations in the Kuiper belt created primarily through capture of objects scattered by Neptune, but also reveals some of the areas where improvement is necessary.
Resonant populations which include the capture of particles through resonance sweeping do not match the parametric distributions of real TNOs from characterized surveys.
However, the outer resonant populations and the scattering, detached, and outer classical populations are an informative comparison sample for real TNOs.
Overall, the B\&M simulation provides a good model of the population sizes and orbital distributions of the distant components of the Kuiper Belt, and provides us with the ability to make the most detailed comparisons between the model and reality to date.  
More simulation work is needed to reproduce the detailed structure in the outer Solar System, including a variety of planet and test particle starting conditions and migration paths, and with the availability of characterized survey results, these simulation results can be statistically tested against real TNO detections.

\bibliographystyle{/Applications/TeX/aastexv6.0/aasjournal}
\bibliography{/Users/repike/Documents/Research/biblio/rpike2015}

\end{document}